\documentclass[prl,aps,showpacs,twocolumn]{revtex4}

\usepackage{graphicx}

\begin{document}
\title{Can Dirac fluid in graphene be made more perfect?}

\author{Sergei Sergeenkov}
\affiliation{Departamento de F\'{\i}sica, CCEN, Universidade Federal da Para\'{\i}%
ba, 58051-970 Jo\~{a}o Pessoa, PB, Brazil}

\date{\today}

\begin{abstract}

To answer this question, we discuss the properties of electronic
viscosity in deformed graphene by introducing strain $\epsilon$
and velocity gradient $\nabla v$ as pseudo-magnetic $B^d \propto
\hbar \epsilon$ and pseudo-electric $E^d \propto \hbar \nabla v$
fields into the Dirac model. We found that kinematic viscosity
$\nu$ depends on applied strain as $\nu \propto \epsilon^{-3/2}$,
decreasing from $\nu\simeq 0.005m^2/s$ at $\epsilon=0.05$ to
$\nu\simeq 10^{-4} m^2/s$ at $\epsilon=0.5$, and simultaneously
leading to a rather significant increase of the Reynolds number
$Re\propto 1/\nu$ from $Re(\epsilon=0.05) \simeq 20$ to
$Re(\epsilon=0.5) \simeq 1000$ which opens a real possibility for
manifestation of noticeable turbulent effects in strained
graphene.
\end{abstract}

\pacs{72.80.Vp, 47.75.+f}

\maketitle

 Still full of surprises, graphene continues to attract attention
as a unique test ground for many interesting ideas from condensed
matter physics, quantum field theory and gravity (see,
e.g.~\cite{i1,i2,i3,i4,i5,3,4,5} and further references therein).
In particular, its electronic properties are shown to closely
follow predictions of the relativistic hydrodynamics with
extremely small value of Dirac fluid viscosity~\cite{a1,a2,a3}.
This phenomenon, coined "an almost perfect electronic fluid", has
been attributed to manifestation of specific topological
properties of graphene. At the same time, many interesting and
unusual phenomena in graphene under mechanical deformations
(leading to substantial modifications of its electronic structure)
have been recently observed or predicted
~\cite{1a,2a,3a,5a,5b,5c,6a,7a,7b,8a}.

In this Letter we discuss the properties of dynamic viscosity of
electronic fluid in a strained graphene by introducing a
homogeneous strain and constant velocity gradient into the Dirac
model via pseudomagnetic and pseudoelectric field, respectively.
The obtained analytical results suggest quite an optimistic
possibility for further improvement of the perfectness of the
Dirac fluid in a strained graphene along with a concomitant
strain-induced change of electronic fluid motion from a laminar to
a turbulent behavior.

Recall~\cite{1a,2a,5c} that in the absence of chirality
(intervalley) mixing, the low-energy electronic properties of
graphene near the Fermi surface can be reasonably described by a
two-component wave function $|\Psi>=(\Psi_A, \Psi_B)$ obeying a
massless Dirac equation
\begin{equation}
{\cal H}|\Psi>=E|\Psi>
\end{equation}
with an effective Hamiltonian
\begin{equation}
{\cal H}=v_F(\sigma_x \pi_x+\sigma_y \pi_y)+eV_d
\end{equation}
Here, $\pi_a=p_a+eA^d_a$ with $p_a=-i\hbar \nabla_a$ being the
momentum operator and $A^d_a=(0,A^d_y)$ the deformation induced
vector potential, $eV_d$ is a chemical potential; $\sigma_a$ are
the Pauli matrices, and $v_F$ is the Fermi velocity. In what
follows, $a =\{x,y\}$. Let us consider viscosity of an electronic
fluid $\eta$ driven by a constant velocity gradient $\nabla v$ in
a square ($L\times L$) graphene sheet under the influence of a
homogeneous strain field $\epsilon$. As is well
known~\cite{1a,2a}, the homogeneous strain $\epsilon$ induced
gauge potential is given by $A^d_y=B^d_zx$ where $B^d_z=\hbar
{\epsilon} /er^2$ is a pseudo-magnetic field inside deformed
graphene with $r=0.14nm$ being carbon-carbon bond length. By
analogy with plastically deformed graphene ~\cite{5c}, we
introduce viscosity effects into the model through a constant
velocity gradient $\nabla_y v$ induced scalar potential
$V_d=E^d_yy$ where $E^d_y=\hbar \nabla_y v/er$ is a
pseudo-electric field created by a moving electronic fluid.
Without losing generality, in what follows we assume that the
velocity $v$ is a scalar and use $\nabla_y v\equiv \nabla v$ to
simplify notations.

It can be easily verified that Dirac equation (1) with the
above-defined potentials have the normalized solutions
\begin{equation}
\Psi_A=C_1\exp\left[i(k_1x+k_2y)-\frac{(x-x_0)^2}{2l_1^2}-\frac{(y+y_0)^2}{2l_2^2}\right]
\end{equation}
and
\begin{equation}
\Psi_B=C_2\left[i\frac{(x-x_0)}{l_1^2}+\frac{(y+y_0)}{l_2^2}\right]\Psi_A
\end{equation}
with $k_1=x_0/l_1^2$, $k_2=y_0/l_2^2$, and the total energy
\begin{equation}
E=\hbar v_F\sqrt{\frac{2}{l_1^{2}}+\frac{2}{l_2^{2}}}
\end{equation}
where $l_1^2=\hbar/eB^d_z$ and $l_2^2=\hbar v_F/eE^d_y$.

Recall that a linear (that is $\nabla v$ independent) coefficient
of dynamic shear viscosity $\eta$ is defined via stress field
$\sigma$ created by a gradient of velocity $v$ as
\begin{equation}
\sigma=\eta \nabla v
\end{equation}
On the other hand, the stress field can be generally defined as a
strain $\epsilon$ induced response of the electronic fluid (within
a graphene sheet of area $A=L\times L$), namely
\begin{equation}
\sigma \equiv -\frac{1}{A}\left [\frac{\partial E}{\partial
\epsilon}\right ]
\end{equation}

As a result, from Eqs.(5)-(7), one obtains
\begin{equation}
\eta \equiv \left [\frac{\partial \sigma}{\partial \nabla v}\right
]_{\nabla v=0}=\left(\frac{\hbar }{L^2} \right)\left(\frac{1
}{2\epsilon} \right)^{3/2}
\end{equation}
for the coefficient of shear viscosity in strained graphene. First
of all, notice that since $\eta \propto \epsilon ^{-3/2}$, the
increase of an applied stress field $\epsilon$ results in
decreasing of the linear viscosity coefficient. More specifically,
in order to compare the predicted here effects with the available
data based on a hydrodynamic approach \cite{a1,a2,a3}, let us
introduce the so-called kinematic viscosity $\nu$ which is related
to the shear coefficient via the mass density $\rho$ of the fluid
as follows, $\nu =\eta /\rho$. According to \cite{a3}, at room
temperature the mass density of electronic fluid in graphene is of
the order of $\rho \simeq 6\times 10^{-19}kg/m^2$ leading to $\nu
\simeq 0.005m^2/s$. It can be easily verified that in our case of
strain induced viscosity, this value of $\nu$ corresponds to
typical values of $\epsilon=0.05$ and $L=1\mu m$. At the same
time, according to our predictions, application of an
experimentally achievable strain \cite{b} of $\epsilon=0.5$ will
result in a drop of kinematic viscosity to $\nu \simeq
10^{-4}m^2/s$. Recall that within the hydrodynamic approach
\cite{a3}, the electron conductivity $\sigma$ is related to $\nu$
as $\sigma \propto 1/\nu$ leading to strain-induced increase of
conductivity $\sigma (\epsilon)\propto \epsilon^{3/2}$ with the
ratio $\sigma (\epsilon=0.5)/\sigma (\epsilon=0.05)\simeq 50$. It
is worthwhile to mention that in graphene electron fluid
velocities can reach as high as \cite{c} $v\simeq 0.1v_F$ leading
to gradients $\nabla v\simeq v/L\simeq 0.1v_F/L\simeq
10^{11}s^{-1}$ for $L=1\mu m$. On the other hand, in addition to a
rather strong decrease of the viscosity, Eq.(8) predicts a real
possibility for manifestation of turbulent like regime in graphene
which is characterized by large enough values of the Reynolds
number $Re=vL/\nu$. Indeed, according to Eq.(8), one obtains
\begin{equation}
Re=\frac{\rho vL}{\eta} =\frac{\rho vL^3(2\epsilon)^{3/2}}{\hbar}
\end{equation}
for strain dependent Reynolds number. Given the above-mentioned
estimates of the model parameters (namely \cite{a1,a2,a3,c},
$\rho=6\times 10^{-19}kg/m^2$, $v=0.1v_F$ and $L=1\mu m$), Eq.(9)
predicts quite a significant increase of the Reynolds number from
$Re\simeq 20$ (at $\epsilon=0.05$) to $Re\simeq 1000$ (at
$\epsilon=0.5$), describing a strain-induced laminar-to-turbulent
transition of electronic fluid in graphene which could be relevant
for potential nanoelectronics applications \cite{a1,d}.

This work has been financially supported by the Brazilian agency
CAPES.


\begin{thebibliography}{99}
\bibitem{i1} A.K. Geim and A.H. MacDonald, Phys. Today {\bf 60},
35 (2007).
\bibitem{i2} A.K. Geim, Science {\bf 324}, 1530 (2009).
\bibitem{i3} K.S. Novoselov, Rev. Mod. Phys. {\bf 83}, 837 (2011).

\bibitem{i4} E. Shuryak, Prog. Part. Nucl. Phys. {\bf 53}, 273 (2004).
\bibitem{i5} P.K. Kovtun, D.T. Son, and A.O. Starinets, Phys. Rev. Lett. {\bf
94}, 111601 (2005).


\bibitem{3} M.I. Katsnelson and  G.E. Volovik, JETP Lett. {\bf
95},  411 (2012).

\bibitem{4} M.I. Katsnelson and  G.E. Volovik, J. Low Temp. Phys. {\bf
175},  655 (2014).



\bibitem{5} G.E. Volovik M.A. Zubkov, Ann. Phys. (NY) {\bf
340},  352 (2014).


\bibitem{a1} M. Muller, J. Schmalian, and L. Fritz, Phys.
Rev. Lett. {\bf 103}, 025301 (2009).

\bibitem{a2} M. Mendoza, B.M. Boghosian, H.J. Herrmann, and S. Succi, Phys. Rev. Lett. {\bf 105}, 014502 (2010).

\bibitem{a3} M.Mendoza, H.J. Herrmann, and S. Succi, Phys. Rev. Lett. {\bf 106},
156601 (2011).






\bibitem{1a} M.A.H. Vozmediano, M.I. Katsnelson, and F. Guinea,
Phys. Rep. {\bf 496}, 109 (2010).

\bibitem{2a} Ken-ichi Sasaki, Riichiro Saito, M.S. Dresselhaus, Katsunori Wakabayashi, and Toshiaki
Enok, New Journal of Phys. {\bf 12}, 103015 (2010).

\bibitem{3a} N. Levy, S.A. Burke, K.L. Meaker, M. Panlasigui, A. Zettl, F. Guinea, A. H. Castro Neto, and M. F. Crommie, Science {\bf 329}, 544
(2010).


\bibitem{5a} S. Sergeenkov and F.M. Araujo-Moreira, Solid State Commun. {\bf 158}, 58  (2013).

\bibitem{5b} S. Sergeenkov and F.M. Araujo-Moreira,
Phys. Lett. A {\bf 377}, 3158 (2013).

\bibitem{5c} S. Sergeenkov and F.M. Araujo-Moreira, JETP Lett. {\bf 98}, 88  (2013).


\bibitem{6a} Jamie H. Warner, Elena Roxana Margine, Masaki Mukai, Alexander W. Robertson,
Feliciano Giustino, and Angus I. Kirkland, Science {\bf 337}, 209
(2012).

\bibitem{7a} Tony Low and  F. Guinea, Nano Letters {\bf 10}, 3551
(2010).

\bibitem{7b} Yongjin Jiang, Tony Low, Kai Chang, Mikhail I. Katsnelson, and Francisco
Guinea, Phys. Rev. Lett. {\bf 110}, 046601 (2013).

\bibitem{8a} Jiong Lu, Yang Bao, Chen Liang Su, and Kian Ping Loh,
ACS Nano {\bf 7}, 8350 (2013).

\bibitem{b} Mark A. Bissett, Masaharu Tsuji, and Hiroki Ago, Phys. Chem. Chem. Phys. {\bf 16},  11124 (2014).

\bibitem{c} Inanc Meric, Melinda Y. Han, Andrea F. Young, Barbaros
Ozyilmaz, Philip Kim, and Kenneth L. Shepard, Nature
Nanotechnology {\bf 3}, 654 (2008).

\bibitem{d} A.K. Saha, K. Muralidhar, and G. Biswas, Journal of Engineering
Mechanics-ASCE {\bf 126}, 523 (2000).


\end{thebibliography}
\end{document}